# Quantitative Ultrasound and B-mode Image Texture Features Correlate with Collagen and Myelin Content in Human Ulnar Nerve Fascicles

Michal Byra[1,2], Lidi Wan[1,2], Jonathan H. Wong[1,2], Jiang Du[1,2],

Sameer B. Shah[1,3], Michael P Andre[1,2], Eric Y Chang[1,2]

[1]Research Service, VA San Diego Healthcare System, San Diego, CA, USA

[2]Department of Radiology, University of California, San Diego, CA, USA

[3]Departments of Orthopedic Surgery and Bioengineering, University of California, San Diego, CA, USA

Corresponding author: Michal Byra, mbyra@ucsd.edu, phone: +1 858 283 9650

Address: Department of Radiology, University of California, San Diego, 9500 Gilman Drive, La Jolla, California, 92093, USA

**Abstract**

We investigate the usefulness of quantitative ultrasound (QUS) and B-mode texture features for characterization of ulnar nerve fascicles. Ultrasound data were acquired from cadaveric specimens using a nominal 30 MHz probe. Next, the nerves were extracted to prepare histology sections. 85 fascicles were matched between the B-mode images and the histology sections. For each fascicle image, we selected an intra-fascicular region of interest. We used histology sections to determine features related to the concentration of collagen and myelin, and ultrasound data to calculate backscatter coefficient (-24.89 dB ± 8.31), attenuation coefficient (0.92 db/cm-MHz ± 0.04), Nakagami parameter (1.01 ± 0.18) and entropy (6.92 ± 0.83), as well as B-mode texture features obtained via the gray level co-occurrence matrix algorithm. Significant Spearman's rank correlations between the combined collagen and myelin concentrations were obtained for the backscatter coefficient (R=-0.68), entropy (R=-0.51), and for several texture features. Our study demonstrates that QUS may potentially provide information on structural components of nerve fascicles.

**Introduction**

Ultrasound (US) is considered a first-line imaging modality for the clinical evaluation of peripheral nerves (Expert Panel on Musculoskeletal, et al. 2015, Expert Panel on Musculoskeletal, et al. 2018). However, important challenges that remain when imaging nerves with clinical scanners include limitations in resolution and reliance on subjective analyses. These limitations are exacerbated by the composite nature of nerve tissue, which comprises well-organized bundles of nerve fibers within a fascicle, and bundles of fascicles within a whole nerve. Commonly-utilized clinical ultrasound transducers for nerve imaging operate with maximal frequencies ranging from 15-22 MHz, with corresponding reported characteristic resolution of approximately 0.1-0.2 mm (Moran, et al. 2011).

To date, nerve characterization is most typically performed using B-mode US images, mainly to subjectively assess morphological features of the whole nerve (Kane, et al. 2004). B-mode images have been used to detect and evaluate various pathologies of the peripheral nerves, for example US can detect complete and partial nerve transections (Lawande, et al. 2014). In the setting of carpal tunnel syndrome, nerves show characteristic morphological features in US images (Cartwright, et al. 2012, Chan, et al. 2011). Ultrasound systems with higher frequencies ranging from 20 to 100 MHz have been introduced into practice, mostly for eye and skin imaging (Lockwood, et al. 1996). The first US scanners operating at frequencies up to 50 MHz were recently cleared by regulatory agencies for human use. These systems permit greater than three-fold improvements in resolution and provide new ways to assess peripheral nerves. These high frequency scanners can be used to provide clear delineation of nerve fascicle boundaries at certain locations, such as the median nerve at the wrist (Cartwright, et al. 2017).

Quantitative ultrasound (QUS) is emerging as an exciting method to potentially reduce reliance on subjective evaluation. Quantification can be employed on routine B-mode images to distinguish between normal and diseased nerves, such as through usage of the density function (fraction of the hypoechoic pixels in a nerve) (Bignotti, et al., 2015, Tagliafico, et al. 2010), or can be used to improve nerve detection, such as through use of Gabor descriptors (Hadjerci, et al. 2014). Using raw radiofrequency (RF) US data, more fundamental ultrasonic parameters can now be quantified in a scanner-independent way, including backscatter coefficient (BSC), attenuation coefficient (AC) and stochastic modeling of envelope statistics of backscattered echoes (Mamou and Oelze, 2013). The BSC is a measure of a tissue's ability to backscatter US waves, providing information that is analogous, but not equal, to qualitative tissue echogenicity dependent on tissue structure and composition. The AC is related to the loss of energy that occurs in tissue during the wave propagation due to absorption and scattering. Modeling of the backscatter echo statistics provides information about the spatial distribution of scattering microstructures within the resolution cell of the imaging transducer.

QUS is now considered an efficient and reliable method of tissue characterization (Oelze and Mamou, 2016) with wide-ranging applications including liver disease assessment (Lin, et al., 2015, Zhou, et al., 2018), breast mass characterization (Byra, et al., 2016, Sadeghi-Naini, et al., 2017) and evaluation of muscular changes (Weng, et al., 2017). QUS methods have also been used with high frequency data for characterization of the annular pulleys of the fingers (Lin, et al., 2017) and skin tissue (Pereyra and Batatia, 2012, Piotrzkowska-Wroblewska, et al., 2015). However, to the best of our knowledge, QUS methodology based on raw RF US data has not been applied to characterize human nerves or its fascicles using high frequencies. A non-invasive method that may potentially provide information on the structural components within nerve fascicles could have wide reaching diagnostic and research applications. Therefore, the purpose of this study was to use US-based features, obtained from a clinical scanner capable of operating at ultra-high frequencies (>20 MHz), for characterization of human ulnar nerve fascicles with histological correlation. Specifically, we employed QUS features calculated from raw RF data, including BSC, AC, Nakagami parameter, and entropy, as well as texture features obtained via the gray level co-occurrence matrix (GLCM) algorithm applied to B-mode images.

**Materials and Methods**

*Sample Preparation*

This prospective study was approved by our Institutional Review Board with a waiver of informed consent for cadaveric specimen work. Six fresh-frozen cadaveric upper extremities with intact shoulder joints (three females, three males; mean age, 48.8 years old; range, 21-70 years) were obtained from a nonprofit whole-body donation company (United Tissue Network, Phoenix, AZ) and included in this study. The specimens were disarticulated at the scapulothoracic and sternoclavicular articulations proximally and cut with a bandsaw at the mid forearm distally. Specimens were stored in an ultra-low freezer (-80 °C) and thawed in a room-temperature water bath for 24 hours prior to scanning. Specimens underwent a single freeze-thaw cycle in total.

*Data Acquisition*

Scanning was performed by a fellowship-trained musculoskeletal radiologist (8 years of experience with musculoskeletal US and radiological-pathological imaging correlation) using a clinical ultrahigh frequency US scanner (Vevo MD, FUJIFILM, Toronto, Canada), a linear probe with nominal 30 MHz center frequency (UHF48 transducer) and standard coupling gel. The sampling frequency was equal to 240 MHz The raw beam-formed RF data were acquired using a constant time gain compensation, and the US signals were amplified with the same factor, which is independent of depth. Imaging was performed with the elbow in approximately 45° flexion and each ulnar nerve was scanned at 9 distinct locations, 5 mm apart, centered at the level of the medial epicondyle, see Fig 1a). Identical scanner settings were used during the data acquisition at each point. Following scanning of each nerve, US raw data were acquired from a custom reference phantom (E. Madsen, University of Wisconsin, Madison, WI), with homogeneous acoustic properties calibrated *a priori* over the range of frequencies encountered in this study. This additional phantom data acquisition was necessary for the calculation of the QUS parameters (Yao, et al. 1990) to normalize scanner-dependent conditions (beam focus, field of view, frequency, gain, etc.).

*Histologic Preparation*

After US scanning, the specimens were carefully dissected, see Fig. 1b). The roof of the cubital tunnel was released and the epineurium of each ulnar nerve was labeled with a permanent marker while still in situ to facilitate accurate matching between the locations of the US image and the histology sections. The ulnar nerves were removed, fixed in 10% zinc formalin (Z-fix, Anatech LTD, Battle Creek, MI), treated with 30% sucrose for cryoprotection, and cut to 6 µm thick cross-sections. Staining with Masson's trichrome (Richard-Allen Scientific, Kalamazoo, MI) was performed with incubation times of 10 minutes for Weigert's hematoxylin, 3 minutes for Biebrich scarlett-acid fuschin, 5 minutes for phosphotungstic (PTA) and phosphomolybdic acid (PMA), and 5 minutes for aniline blue, to identify connective tissue (blue component), and myelinated axons (violet-red), which had a color distinct from the bright red of non-basophilic structures typically associated with trichrome staining.

*Data Analysis*

54 pairs of histology and B-mode images were acquired (6 specimens, each with 9 spatially distinct imaging locations). The same musculoskeletal radiologist who performed the US scanning reviewed the pairs of images and matched the fascicles on the histology images to those on the B-mode images, Fig. 1c)-e), and outlined the regions interest (ROIs) indicating the fascicles. With regards to Fig. 1, we note that the selected B-mode image is a single representative still image. The experienced radiologist had the benefit of real-time scanning, and at multiple locations in this specimen, and annotated images immediately after scanning to guide future analysis. Previous authors have similarly found that real-time scanning leads to improved structural assessment (Clair, et al. 1982, Van Holsbeke, et al. 2008). Fascicles that were >1 mm in diameter were first identified on both the US and histology images. Based on these larger fascicles, histology images were either flipped or rotated to match the US images. Thereafter, attempts were made for inter-modality fascicle correlation based on size, spatial relationship, and orientation relative to these larger fascicles. Only fascicles that could be well-delineated for at least 50% of its perimeter were included. Specifically, fascicles were not included when the majority of the interface at the internal epineurial boundary was poor on the US images or the fascicles could not be confidently matched between modalities by an experienced musculoskeletal radiologist. An ultrasound medical physicist agreed upon the selected ROIs and confirmed the correspondence of fascicles. Moreover, matched fascicles were excluded from analysis when histology images contained obvious processing or sectioning artifacts, such as folds or tissue damage/separation. In total, 85 fascicles out of 29 image pairs were included. From the six specimens we extracted 3, 9, 11, 14, 20 and 28 pairs for the analysis, respectively.

Histology images were segmented using a color thresholding method (Fig. 2). First, the original RGB (red-green-blue) images were used to extract the white spaces. Second, each image was transformed from the RGB to the YUV (luminance-bandwidth-chrominance) color space to more easily separate the blue from the purple areas, indicating collagen (CL) and myelin (ML), respectively (Huisman, et al. 2015, Turedi, et al. 2018). In the case of the V chrominance component of the YUV color space, blue corresponds to the opposite part of the color spectrum than purple. Following segmentation, the area fractions of regions positive for CL and ML were calculated. In addition, the ML to CL ratio and ML and CL sum (MCL) were calculated. All calculations were performed in Matlab (MathWorks, 2018a, Natick, MA).

*Quantitative Ultrasound Features*

Attenuation Coefficient

Due to the tissue attenuation, US wave amplitude decreases with the propagation distance. The frequency-dependent attenuation is usually modeled using the following equation:

$$A(z) = A(z_0) - \alpha f z, \tag{1}$$

where $A(z)$ is the echo amplitude (in dB) decreased from the initial amplitude $A(z_0)$ at point $z_0$ over the propagation distance $z$, $f$ stands for the frequency, $\alpha$ (in dB/cm-MHz) is the AC. The AC can be estimated using the reference phantom with the formula:

$$AC(f) = AC_{phantom}(f) - \frac{\gamma(f)}{4 * 8.686}, \tag{2}$$

where $AC_{phantom}(f)$ is the pre-computed reference phantom AC and $\gamma(f)$ is the slope of the straight line that fits the log spectrum difference of the backscattered echoes in the tissue of interest and the reference phantom as a function of depth (Han, et al. 2017).

Backscatter Coefficient

The BSC quantifies the ability of the tissue to backscatter ultrasound energy and provides a measure of tissue "echogenicity." Similar to the AC, the BSC depends on the frequency of the imaging pulse, because the backscattering occurs on tissue microstructures that are small in comparison to the wavelength (Mamou and Oelze 2013) . The BSC can be calculated using the reference phantom method following attenuation compensation using the equation (Han, et al. 2017) :

$$BSC(f) = \frac{S(f)}{S_{phantom}(f)} BSC_{phantom}(f) 10^{0.2(AC(f) - AC_{phantom}(f))}, \tag{3}$$

where $S(f)$ and $S_{phantom}(f)$ are the sample and the reference phantom power spectrum functions, respectively. $BSC_{phantom}(f)$ is the pre-computed BSC for the reference phantom, and the exponential factor is present in the equation due to the attenuation correction (Yao, et al. 1990). The effects of beamforming and gain are compensated using the reference phantom method.

Envelope Statistics

The Nakagami distribution was used to assess the envelope statistics of backscattered echoes in each nerve fascicle. The probability density function of the Nakagami distribution can be expressed as follows (Mohana Shankar 2000) :

$$f_N(A) = \frac{2m^m A^{2m-1}}{\Gamma(m)\Omega^m} \exp\left(-\frac{m}{\Omega} A^2\right) \qquad (4)$$

where $A$ represents the backscattered echo amplitude, $\Gamma(\cdot)$ is the Gamma function, $m$ and $\Omega$ are the shape and the scaling parameters of the Nakagami distribution, respectively. The Nakagami parameter $m$ is sensitive to the spatial distribution of tissue micro-structures within the resolution cell of the imaging transducer. For $0.5 < m < 1$ the Nakagami distribution is pre-Rayleigh, interpreted as meaning the resolution cell contains a small number of strong scatters mixed with the randomly distributed scatterers. For $m = 1$ the Nakagami distribution is identical to the Rayleigh distribution and the resolution cell is expected to contain a large number of randomly distributed scatterers. For $m > 1$ the Nakagami distribution becomes Rician; the resolution cell contains a large number of randomly distributed scatterers along with regularly spaced scatters.

Additionally the Shannon entropy of the backscattered echo amplitude was used for nerve characterization (Zimmer, et al. 1996). In comparison to the Nakagami imaging, this is a different, model-free, approach to the analysis of the backscattered echo statistics and has been recently applied for liver fat assessment (Tsui and Wan 2016, Zhou, et al. 2018). The Shannon continuous entropy can be expressed in the following way:

$$\text{Entropy} = -\int_0^\infty f(A)\ln(f(A))\,dA, \qquad (5)$$

where *A* is the US backscattered echo amplitude and *f(A)* refers to the probability density function of *A*. Entropy is also strictly related to the Nakagami distribution, for which entropy can be expressed in the following way (Zimmer, et al. 1996):

$$\text{Entropy}_{\text{Nakagami}} = \frac{1}{2}\ln\Omega + h(m), \tag{6}$$

where $\Omega$ and *m* are the parameters of the Nakagami distribution, $h(\cdot)$ is a function of the Nakagami parameter. Equation (5) clearly depicts that entropy is sensitive to the scaling of echo amplitudes (due to the time-gain compensation, for example) and requires attenuation correction, similar to the BSC, to be quantitative. Following the attenuation correction, it provides a parameter that jointly describes tissue echogenicity and backscattered envelope statistics.

Image Texture Analysis

Image texture analysis was performed using the GLCM algorithm (Haralick, et al. 1973), a frequently employed technique in US imaging (Andrade, et al. 2012, Bharti, et al. 2016, Flores, et al. 2015, Gaitini, et al. 2004, Sadeghi-Naini, et al. 2017). GLCM contains co-occurrence probabilities of all pair wise combinations of grey levels in an image or image region. Given a GLCM, the following six texture features were extracted: contrast, correlation, entropy, homogeneity, energy, maximum probability (Clausi 2002). These features quantify the fundamental properties of image texture related to its uniformity or how the pixels in the image are correlated with each other.

*US Feature Calculation*

The feature estimation flowchart is depicted in Fig. 2. First, the frequency range of the backscattered echoes from the fascicles was determined using the Fourier transform as knowledge of the range of frequencies is important for QUS parameter calculation. The average center frequency of the backscattered echoes was determined to be approximately 16 MHz.

The downshifted center frequency was expected for these high frequencies, caused by the attenuation of the US wave in the tissue layers superficial to the nerve (Klimonda, et al. 2016). The BSC and the AC were estimated for the frequency range as follows. Erosion operation was applied to clip each fascicle ROI and remove possible edge effects on the US parameter estimation. The ROIs were eroded with a circular window of size $\lambda/2$, where $\lambda$ is the wavelength in water at 16 MHz. In the next step, each processed ROI was used to fit a square ROI. Square ROIs were used to keep the same number of samples in the RF signal lines oriented with the US A-lines. Next, to compensate attenuation effects in the nerve, the following strategy was employed. First, a uniform region of the tissue superficial to the nerve was selected and used to calculate the AC with the reference phantom method. Second, the AC in each fascicle using the square ROI and the reference phantom method was calculated. Next, both ACs were used to compensate for the effect of the attenuation in each fascicle ROI. The $\gamma(f)$ in eq. 2 was determined using tissue and reference spectrum difference calculated for the frequency range of 14-18 MHz (central portion of the backscattered spectrum). Given the $\gamma(f)$ and the reference phantom AC, using eq. 2 the fascicle ACs were computed for the frequency range of 14-18 MHz and the averaged AC values over the range of 15-17 MHz were used as the estimate of the AC for each fascicle (Han, et al. 2017). Following the attenuation compensation, the BSC was estimated using the reference phantom method. As in the case of the AC, the BSC was calculated for the frequencies between 14 MHz and 18 MHz. Next, to obtain a single estimate, the BSC was averaged over the range of 15-17 MHz. To calculate the envelope statistics parameters, the Hilbert transform was applied to compute RF signal amplitude, and the amplitude samples were extracted using the eroded fascicle ROI. The Nakagami parameter was calculated using the maximum likelihood estimator (Lin, et al. 2017) . Entropy was calculated using eq. 5.

There are two parameters associated with the GLCM algorithm, the distance path (inter-pixel distance) and the orientation angle. In the case of the US B-mode images, the distance path was set to be equal to $\lambda/4$ and the orientation angle was set to 90°. Only the co-occurrence of the gray level values in the axial direction were analyzed.

*Statistics*

Descriptive statistics were calculated, including mean, standard deviation (sd), and coefficient of variation (CV). Correlations between US and histology features as well as between various US features were performed using Spearman's rank. In addition, since adjustment of threshold levels of the B-mode images is commonly used in clinical practice to enhance contrast, the effects of B-mode image reconstruction on the performance of the GLCM-based features were also evaluated. The B-mode images were reconstructed from the amplitude samples, which were logarithmically compressed using following equation:

$$A_{\log} = 20 log_{10}\left(\frac{A}{A_{\max}}\right), \qquad (7)$$

where $A$ and $A_{log}$ are the amplitude and the logarithmically compressed amplitude of the ultrasonic RF signal, respectively. $A_{max}$ indicates the highest value of the amplitude found in the RF data. For a specific threshold level, the log-compressed amplitude was mapped to the range of [0, 255] (8 bits). B-mode images were reconstructed using threshold levels of [30, 35, 40, 45, 50, 55, 60] dB, which are commonly employed in US imaging. Next, we investigated the impact of the reconstruction threshold level on the correlation between the B-mode based features and the histology findings.

**Results**

The average fascicle area, measured using histology images, was equal to 0.81 mm ± 0.59. The estimated QUS parameters for the data acquired from all nerve fascicles are shown Table 1. The smallest CV value was obtained for the AC (CV = 0.044), suggesting that the fascicles attenuated the US waves comparably. The highest CV value was obtained for the BSC (CV =

0.33). The high dynamic range of the BSC (CV almost ten times more than the AC) suggests high variability in the biological/structural contribution to this measure.

The Spearman's rank correlation results between US and histology features are depicted in Table 2. Changes of US parameters were mainly related to the variation of collagen in the fascicles. The highest correlation coefficient ($\rho$ = -0.68) was obtained for the BSC and the MCL. Negative correlation coefficients indicate that the BSC decreases with the MCL (Fig. 3). Entropy demonstrated a lower, but still statistically significant coefficient ($\rho$ = -0.51). Statistically significant correlation coefficients were not found for the Nakagami parameter or the AC. There was also no significant correlation between the fascicular areas determined in histology and the features related to the collagen and myelin concentration.

The Spearman's rank correlation results between various US features are depicted in Table 3 and Fig. 4, which show that the GLCM based features are overall highly correlated with each other. Additionally, the entropy and the BSC are correlated, which is expected since both assess tissue echogenicity. Additionally, high correlation coefficients between the BSC, entropy and the GLCM based features were obtained. As would be expected, QUS entropy was highly correlated with the GLCM entropy, and a high value of the BSC corresponded to a high value of the B-mode image texture homogeneity.

Evaluation of the B-mode image reconstruction threshold level on performance of GLCM-based features demonstrated that the best performance was obtained for the threshold level of around 45 dB. Fig. 5 shows how the Spearman's correlation coefficient between the homogeneity feature and the MCL is modified due to the modification of the threshold level. For high threshold values, the image was dominated by noise and the GLCM feature poorly correlated with histology. Similarly, small threshold values resulted in lower performance due to compression of speckle patterns that corresponded to lower amplitude echoes. These small amplitude values lost their image conspicuity. For analysis used in the remainder of the study, GLCM based features were calculated using the threshold level of 45 dB. As in the case of the QUS parameters, the GLCM based features were related to the CL and the MCL (Table 2). The highest correlation coefficient was obtained for the homogeneity feature, which measures the uniformity of image texture, and the MCL (Fig. 6, $\rho = 0.64$).

**Discussion**

In our study, we used US-based features for ulnar nerve characterization. We imaged the nerve *in situ* at room temperature within the cubital tunnel of cadaveric elbow specimens and successfully matched a large number of the fascicles on the B-mode and histology images. For the matched fascicles, several QUS parameters correlated with histological outcomes, with the highest correlation coefficient of -0.68 for the BSC versus the MCL. Table 2 depicts that the US parameters were mostly related to the collagen variations calculated using the segmented histology images. Negative correlation coefficient between the BSC and the CL suggests that high connective tissue concentration in the fascicles results in a low value of the BSC, or lower overall tissue echogenicity. Backscattering of ultrasound waves occurs at interfaces of tissue microstructures that have different physical properties (e.g. sound speed, density) (Mamou and Oelze, 2013). In a material that has uniform spatial distribution of the physical properties (e.g., local density) the scattering would be low. Our results may agree with this observation; as the CL decreases, the fascicle composition tends to be less uniform, with a possibly larger number of different tissue interfaces that contribute to the backscattering. Moreover, the correlation coefficient between the BSC and the MCL was higher than for the BSC and the CL, which suggests that the myelin concentration also has impact on the backscattering. The AC did not correlate with the histology findings. The average value of the AC was equal to 0.92 dB/cm-MHz, being higher than the average value of the AC in human soft tissue, 0.54 dB/cm-MHz, or the brain, 0.6 dB/cm-MHz (Culjat, et al., 2010, Mast 2000), and highly invariable across fascicles (low CV). The Nakagami parameter also did not correlate well with the histology findings. This may be explained by the average Nakagami parameter of the fascicles, which was close to 1 (Rayleigh scattering), indicating that the investigated tissue contains a large number of randomly distributed scatterers. The Nakagami parameter mostly provides information about the relative scattering scenario occurring in a particular tissue rather than a quantitative assessment of tissue microstructure.

The correlation coefficients obtained for the GLCM based features were slightly lower than for the BSC, with the highest coefficient equal to 0.64 for the GCLM homogeneity feature. The GLCM features, however, were highly correlated with the BSC and the entropy suggesting that these features, as expected, are related to the backscattering properties of the tissue. The estimation of the GLCM features is more scanner dependent than the QUS features. Our results regarding the modification of the threshold levels clearly depicted this issue (Fig. 5). For researchers interested in tissue characterization, QUS parameters would be recommended as the preferential first-line tool. However, if RF data acquisition is not possible, either due to limitations of the scanner, the GLCM based approaches should be also useful. In this scenario, investigators should be aware of the pitfalls and limitations of B-mode image texture analysis and attempt to maintain identical scanner settings and tissue geometry during measurements in order to compensate for the effects of attenuation.

Several studies have previously shown a relationship between tissue echogenicity and collagen concentration, including for native tissues such as the intestines (Kimmey, et al., 1989), and engineered tissues (Deng, et al., 2016, Mercado, et al., 2015). To the best of our knowledge, ours is the first study to utilize QUS methods to assess the physical properties of nerve fascicles. The correlation between the US features and the fascicle collagen and myelin content determined in our study may be useful for the non-invasive assessment of peripheral nerve composition and changes. Importantly, our results showed that this information may not be simply determined by measuring cross-sectional area. Similar to our results, one study found that morphological determinations of cross-sectional area along the ulnar nerve did not correlate with quantitative collagen data (McFarlane, et al,. 1980). Assessment of collagen can be important since it varies widely depending on nerve location (McFarlane, et al., 1980) and/or its physiological or biomechanical environment. For example, nerve immobilization results in myelin degeneration and deposition of collagen in the endoneurium (Culjat, et al., 2010).

Our results complement those from several other investigators who have suggested that QUS can be used to characterize nerves. The AC parameter of the whole rat sciatic nerve in vitro was used to estimate nerve damage due to diabetic peripheral neuropathy (Chen, et al. 2014). Tissue echogenicity was assessed via the nerve density measure and was found to be an important factor in diagnosis of carpal tunnel syndrome (Tagliafico, et al. 2010). In future work, we plan to compare B-mode measures with QUS measures to assess diagnostic performance. QUS methodology can also be used to monitor nerve regeneration (Schmidt and Leach 2003, Wood, et al. 2011) and to measure temperature changes in nerve during high intensity focus ultrasound treatment (Foley, et al. 2008, Lewis, et al. 2015).

There are several limitations to our study. First, the US data was acquired from cadaveric specimens that underwent a freeze-thaw cycle, likely impacting imaging outcomes and histological quality. Caution should be exercised when comparing our results to *in vivo* data. Second, following the US data acquisition, the specimens were dissected to extract the ulnar nerve in preparation for histology. While this was carefully performed, the histology images may not correspond exactly to the imaging plane of the US transducer or represent any distortion in geometry imposed by surrounding anatomical structures *in situ*. Third, outlining of the same fascicles on the B-mode and histology images was performed by an experienced radiologist, but for several B-mode images the visibility of the fascicles was too poor to delineate the fascicles with high confidence. Of note, an *in vivo* study utilizing a higher frequency transducer (nominal center frequency 50 MHz) on the median nerve at a superficial location noted similar findings where some fascicles were easier to depict than others (Cartwright, et al., 2017). One explanation for this may be that the ulnar nerves were imaged with the elbows in approximately 45° flexion. At submaximal tensions, especially in the highly compliant joint region of the ulnar nerve (Mahan et al., 2015), there may be intra-neural fascicular undulations or bending (Shah, 2017), resulting in an oblique imaging plane or potential volume averaging. Fourth, our modeling of the nerve fascicle was incomplete as evidenced by the inability for the variability of BSC, or any other US feature employed in our study, to be fully explained by variability in the connective tissue and myelin content. Indeed, modeling of tissue is challenging, since there are several factors that contribute to backscattering and speckle patterns. We did not investigate the effects of tissue microstructures orientation and position on backscattering (Han and O'Brien 2015), we did not evaluate collagen microarchitecture in three-dimensions, and we only used information from one histologic stain. It is possible that additional nerve constituents which may have been highlighted by different histologic preparations or immuno-stains may have further improved the correlations. Fifth, we limited our analysis to intrafascicular regions, given the ambiguity of perineurial and epineurial boundaries in US images. Therefore, the role of extra-fascicular collagen on QUS outcomes remains an interesting and important area of future work. Finally, though a medical physicist agreed upon ROI selection and confirmed fascicular correspondence between B-mode and histological images, and though there is typically strong reproducibility in nerve morphology and echogenicity measurements among different operators (Cartwright, et al. 2017, Tagliafico, et al., 2010) our study only included one musculoskeletal radiologist for scanning and fascicle correlation. Therefore, we were unable to assess the reproducibility of US measurements for different operators, scanners or transducers. Future work will investigate these variables and also consider employing automatic fascicle segmentation algorithms.

In this work we investigated the usefulness of two US based approaches for nerve characterization. The first one utilized QUS methods and the second one was related to the B-mode texture analysis with the GLCM. Our study shows that both approaches may be useful for nerve characterization, as both yielded significant correlation with histological CL and MCL. However, the usefulness of the GLCM based features depended on the applied B-mode image reconstruction algorithm. Our study suggests that QUS methods, which are scanner independent, should be favored. If access to raw US data is limited, the GLCM based approach still can provide useful information about the investigated tissue.

**AKNOWLEDGEMENT**

The authors acknowledge grant support from the VA Rehabilitation Research & Development Service (I21RX002367, I01RX002604, I01RX001471) and VA Clinical Science Research & Development Service (I01CX001388).

**Conflict of interest statement**

The authors do not have any conflicts of interest.

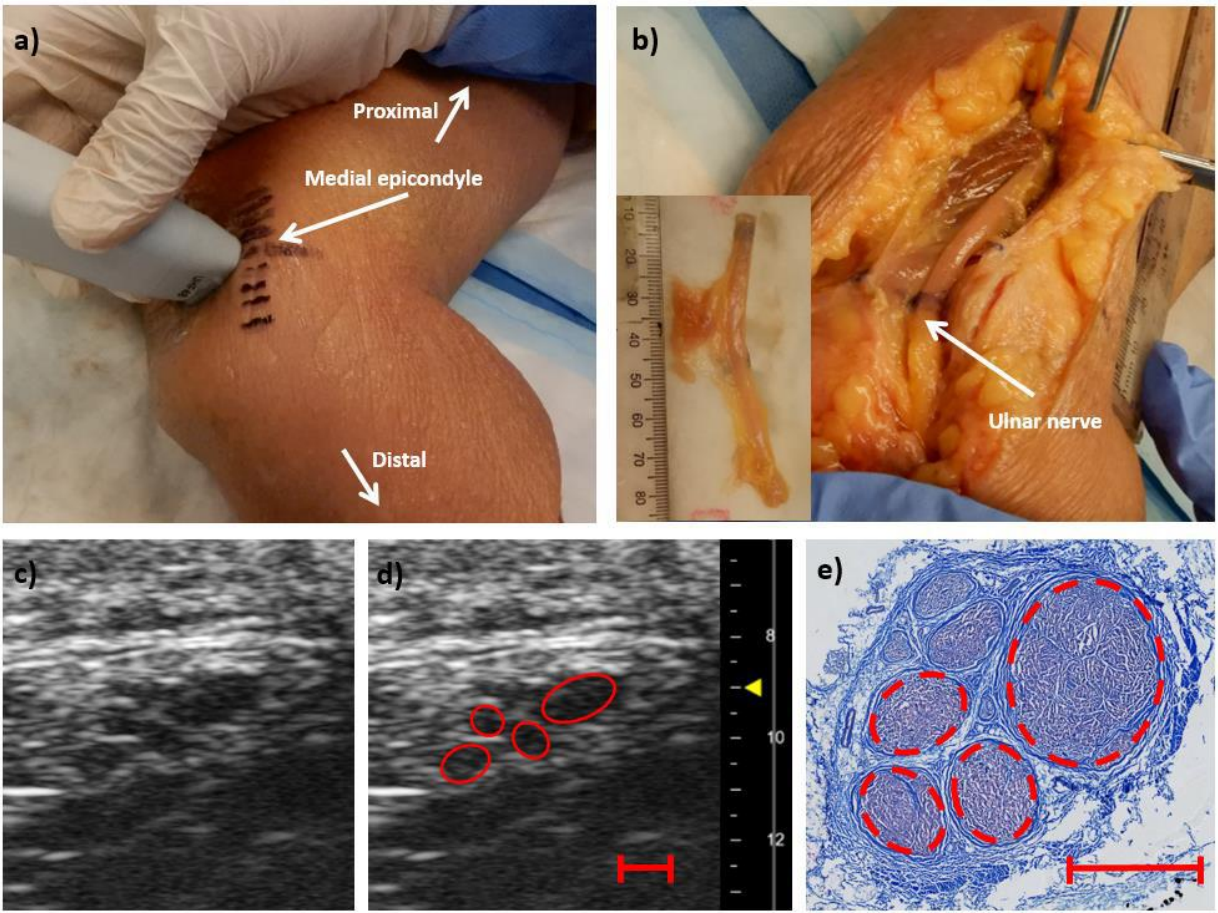

Figure 1: a) Each ulnar nerve was scanned at 9 distinct locations, centered at the level of the medial epicondyle. b) After the scanning the cubital tunnel was released and nerve was removed. The same fascicles on the B-mode image, c) unlabeled and d) labeled, and e) the histology image were selected. Scale bar in mm.

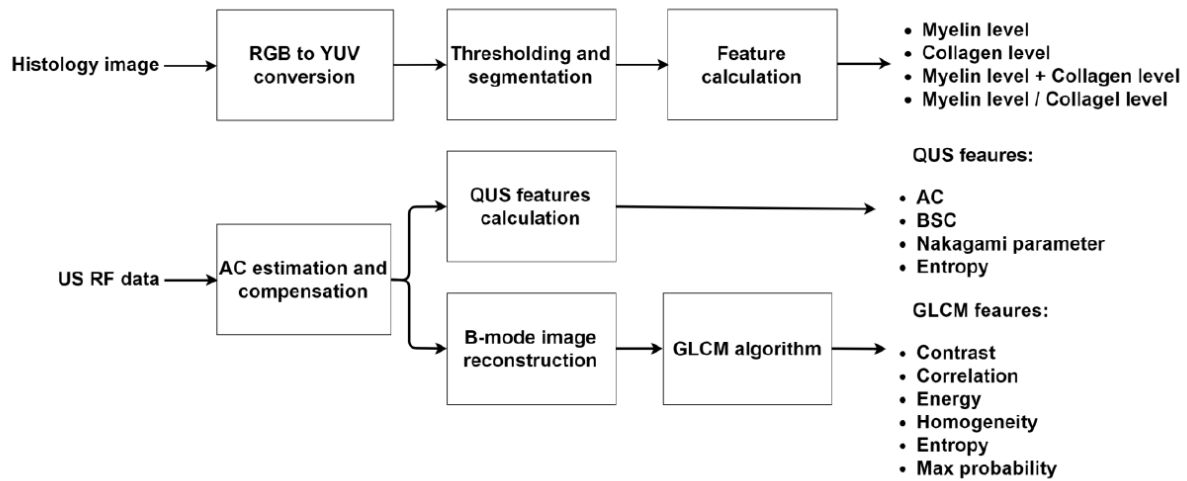

Figure 2: Flowcharts illustrate histology and ultrasound feature extraction from nerve fascicles. The RGB histology images were transformed to YUV color space and color thresholded to extract pixels corresponding to myelin and collagen. Raw ultrasound RF data were used to calculate quantitative parameters, and in the next step employed to reconstruct B-mode images and estimate GLCM (gray level co-occurrence matrix) features.

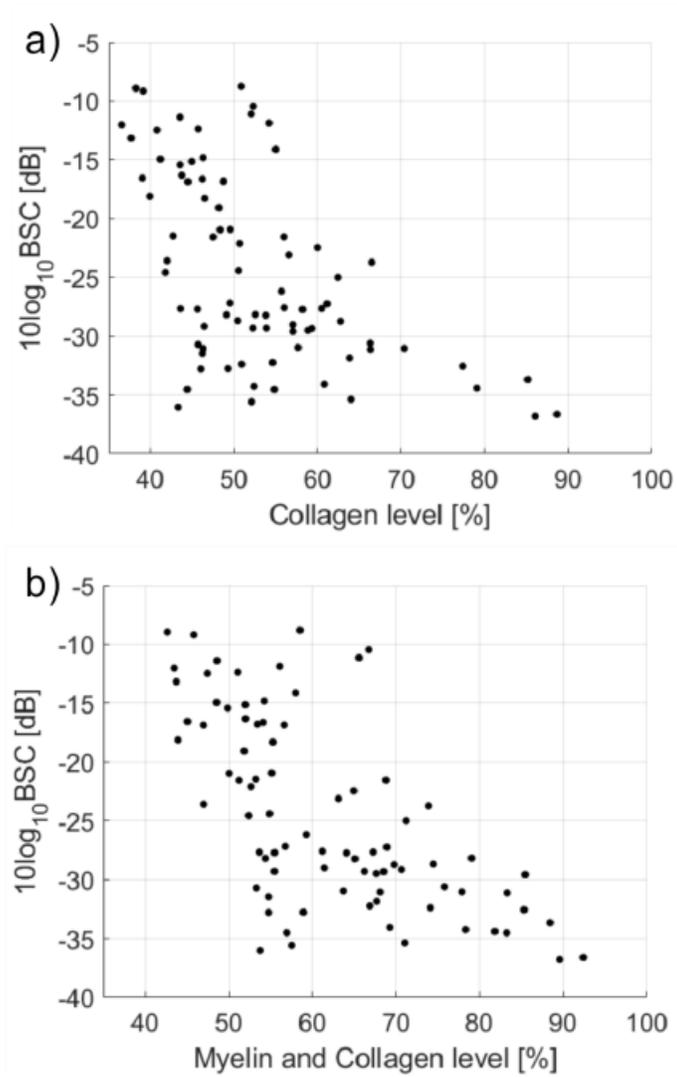

Figure 3: The relation between the BSC (backscatter coefficient) and a) the CL (collagen level), correlation coefficient of -0.56 and b) MCL (myelin and collagen level), correlation coefficient of -0.68.

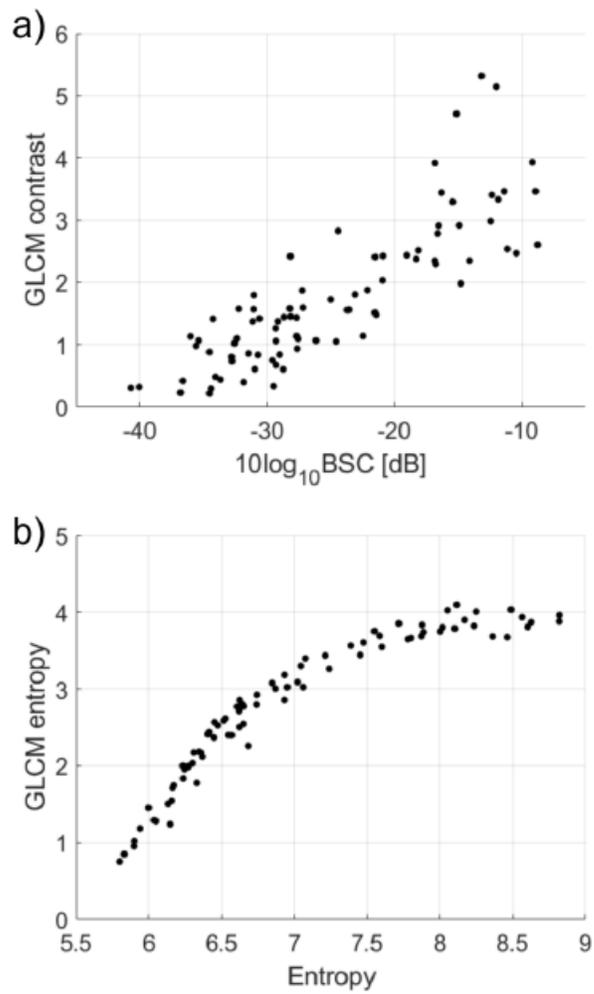

Figure 4: The relation between the QUS (quantitative ultrasound) parameters and the GLCM (gray level co-occurrence matrix) based features. a) For the BSC (backscatter coefficient) the highest correlation coefficient was obtained for the GLCM contrast, 0.87, b) the relation between the QUS entropy and the GLCM entropy, correlation coefficient of 0.98.

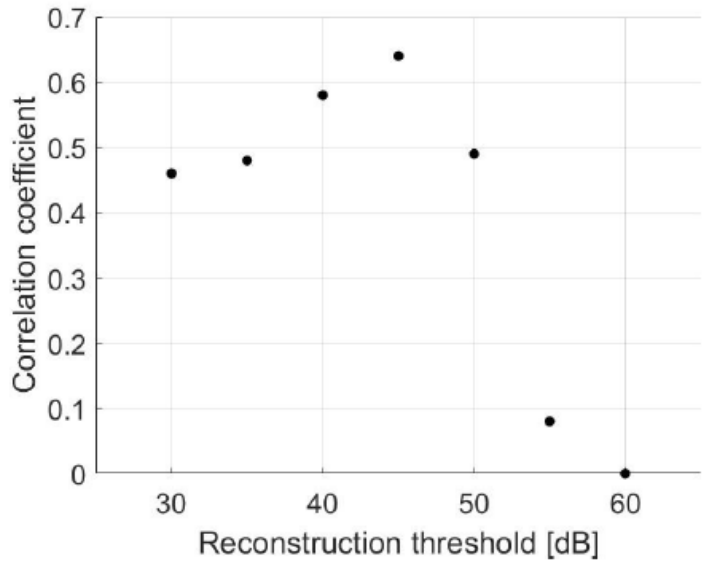

Figure 5: Impact of B-mode image reconstruction threshold level on correlation coefficient between the GLCM (gray-level co-occurrence matrix) homogeneity and the MCL (myelin and collagen level). Based on this plot, 45 dB reconstruction threshold was used for subsequent analysis.

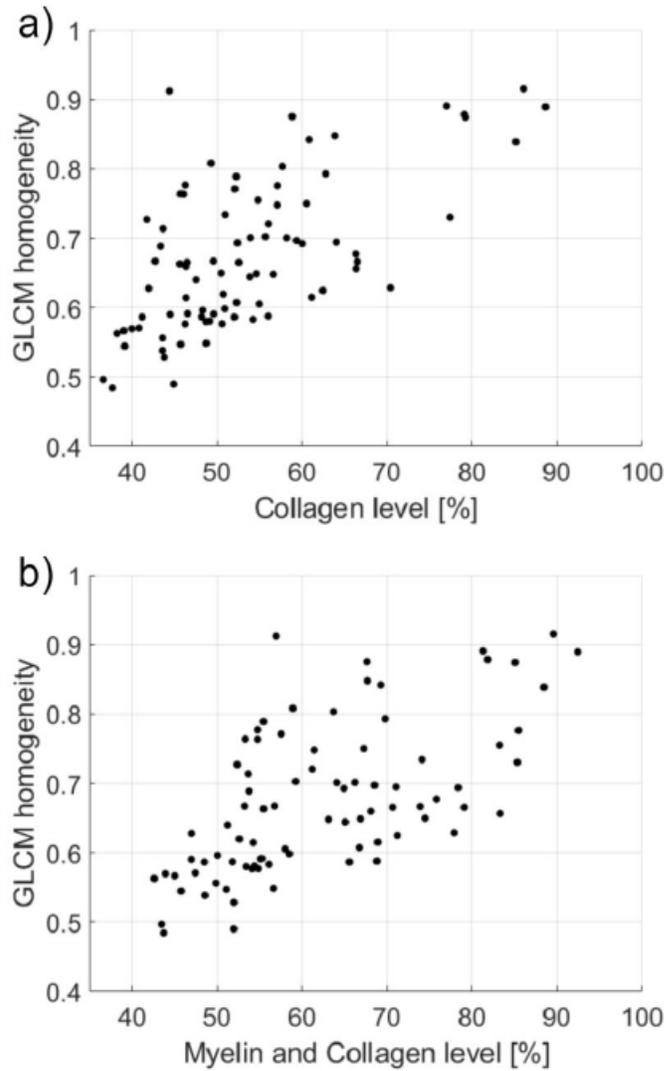

Figure 6: The relation between the GLCM (gray-level co-occurrence matrix) homogeneity and a) the CL (collagen level), correlation coefficient of 0.59 and b) MCL (myelin and collagen level), correlation coefficient of 0.64.

**Tables**

Table 1: Average values of the quantitative ultrasound parameters estimated for the data acquired from 85 fascicles. The AC and BSC refer to the attenuation coefficient and backscatter coefficient, respectively.

| Parameter | AC [dB/cm-MHz] | BSC [dB] | Nakagami | Entropy |
|---|---|---|---|---|
| Mean±sd | 0.92±0.04 | -24.89±8.31 | 1.01±0.18 | 6.92±0.83 |
| CV (sd/mean) | 0.044 | 0.33 | 0.18 | 0.11 |

Table 2: Spearman's rank correlation coefficients between the ultrasound features and the histology findings. Asterisks indicate statistically significant result ($p$-value<0.001).

| Parameter | ML/CL | ML | CL | MCL |
|---|---|---|---|---|
| AC | 0.01 | -0.05 | -0.06 | -0.05 |
| BSC | 0.01 | -0.20 | -0.56* | -0.68* |
| Nakagami | -0.08 | -0.11 | -0.021 | -0.09 |
| Entropy | 0.09 | -0.07 | -0.47* | -0.51* |
| GLCM contrast | 0.11 | -0.08 | -0.58* | -0.63* |
| GLCM correlation | -0.09 | -0.02 | 0.16 | 0.15 |
| GLCM energy | -0.09 | 0.09 | 0.55* | 0.59* |
| GLCM homogeneity | -0.09 | 0.01 | 0.59* | 0.64* |
| GLCM entropy | 0.10 | -0.07 | -0.52* | -0.55* |
| GLCM max prob | -0.05 | 0.13 | 0.56* | 0.62* |

Table3. Spearman's rank correlation matrix for the ultrasound features. The AC, BSC and GLCM stand for the attenuation coefficient, backscatter coefficient and gray level co-occurrence matrix, respectively.

| Parameters | AC | BSC | Nakagami | Entropy | GLCM$_{contrast}$ | GLCM$_{correlation}$ | GLCM$_{energy}$ | GLCM$_{homogeneity}$ | GLCM$_{entropy}$ | GLCM$_{maxprob}$ |
|---|---|---|---|---|---|---|---|---|---|---|
| AC | 1 | | | | | | | | | |
| BSC | 0.21 | 1 | | | | | | | | |
| Nakagami | -0.37 | -0.09 | 1 | | | | | | | |
| Entropy | 0.39 | 0.82 | -0.43 | 1 | | | | | | |
| GLCM$_{contrast}$ | 0.34 | 0.87 | -0.28 | 0.94 | 1 | | | | | |
| GLCM$_{correlation}$ | 0.34 | 0.12 | -0.65 | 0.44 | 0.17 | 1 | | | | |
| GLCM$_{energy}$ | -0.41 | -0.83 | 0.32 | -0.95 | -0.95 | -0.31 | 1 | | | |
| GLCM$_{homogeneity}$ | -0.36 | -0.85 | 0.25 | -0.92 | -0.99 | -0.17 | 0.97 | 1 | | |
| GLCM$_{entropy}$ | 0.40 | 0.84 | -0.38 | 0.98 | 0.96 | 0.36 | -0.99 | -0.96 | 1 | |
| GLCM$_{maxprob}$ | -0.40 | -0.83 | -0.26 | -0.92 | -0.94 | -0.28 | 0.98 | 0.96 | -0.97 | 1 |